\documentclass{aims}
  \usepackage{amsmath}
  \usepackage{paralist}
  \usepackage{graphics} 
  \usepackage{epsfig} 
  \usepackage{amsfonts}
  \usepackage{cases}
  \usepackage{multirow}
  \usepackage{caption}
  \usepackage{array}
  \usepackage{algorithm}
  \usepackage{algorithmic}
  \usepackage{hyperref}
  \usepackage{amssymb}
\usepackage{amsfonts}
\usepackage{graphicx}

\def\currentvolume{X}
 \def\currentissue{X}
  \def\currentyear{200X}
   \def\currentmonth{XX}
    \def\ppages{X--XX}

\newtheorem{theorem}{Theorem}[section]

\newtheorem{lemma}[theorem]{Lemma}

\theoremstyle{definition}

\title[Running heading with forty characters or less]
      {Strongly Convex Programming for Principal Component Pursuit}

\author[first-name1 last-name1 and first-name2 last-name2]{}


\begin{document}
\maketitle

\centerline{\scshape Qingshan You and Qun Wan }
\medskip
{\footnotesize
 \centerline{the School of Electronic Engineering}
   \centerline{University of Electronic Science and Technology of China}
     \centerline{Chengdu, Sichuan, 611731, P. R. China}
} 

\medskip

\centerline{\scshape Yipeng Liu }
\medskip
{\footnotesize
 \centerline{ Department of Electrical Engineering, ESAT-SCD / IBBT - KU Leuven Future Health Department, KU Leuven}
   \centerline{Kasteelpark Arenberg 10, box 2446, 3001 Heverlee, Belgium}
} %

\bigskip

 \centerline{(Communicated by the associate editor name)}

\begin{abstract}
In this paper, we address strongly convex programming for principal component pursuit with reduced linear measurements, which decomposes a superposition of a low-rank matrix and a sparse matrix from a small set of linear measurements. We first provide sufficient conditions under which the strongly convex models lead to the exact low-rank and sparse matrix recovery; Second, we also give suggestions on how to choose suitable parameters in practical algorithms.
\end{abstract}

\section{Introduction}
Recently, much attention has been drawn to the problem of recovering a target matrix from a small set of linear measurements. The estimated matrix is a superposition of low-complexity structure. It can be found in many different fields, such as medical imaging \cite{J10,JA09,A97}, seismology \cite{J73}, information retrieval \cite{C00} and machine learning \cite{A07}.

This problem regained great attention after the publication of the pioneering works of E.J. Cand\'es et al \cite{M02,EB09,EY10,ET10}. According to paper \cite{EX09}, we can build the data model as follows: there exists a large-scale data matrix $ M=L_0+S_0$, where $L_0\in\mathbb R^{n\times n}$ has low-rank, and $S_0$ is sparse component. The main question is how to recover a low-rank matrix $L_0$ and sparse matrix $S_0$ from a small set of linear measurements. In the paper \cite{EX09}, E.J. Cand\'es et.al proved that most low-rank matrices and the sparse components can be recovered, provided that the rank of the low-rank component is not too large, and the sparse component is reasonably sparse; and more importantly they proved that it can be done by solving a simple convex optimization problem, i.e. most matrices of low-rank and the sparse components can be perfectly recovered by solving the optimization problem
\begin{eqnarray}
\mbox{minimize}
& &\| L\|_* + \lambda \| S\|_1   \nonumber \\
\mbox{subject to} & & L+S = M
\end{eqnarray}
provided that the rank of the matrix $L$ and the cardinality of the sparse component $S$ obey
\begin{eqnarray}
\mbox{rank}(L_0)\leq \rho_rn\mu^{-1}(\log{n})^{-2} ~~\mbox{and} ~~m\leq \rho_sn^2 \nonumber
\end{eqnarray}
where  $\rho_r$ and $\rho_s$ are positive numerical constants.

 In practice, it is necessary to develop efficient and effective tools to process, analyze, and extract useful information from such high dimensional data (in application, dimensional of data is always very high). Because strongly convex optimizations have many advantages, such as optimal solution is unique, many scholars suggest solving their strongly convex approximations, see, e.g., \cite{J09,JAS,J08,Z12}, instead of directly solving the original convex optimizations. Pertaining to problem (1), the authors gave the suitable sufficient conditions under which the strongly convex models lead to the exact low-rank and sparse matrix recovery.  Some suggestions were given on how to choose suitable parameters in practical algorithms in the paper \cite{Z12}. However, the results of paper \cite{Z12} are limited in a special case, i.e. $Q=\mathbb R^{n\times n}$. In this paper, we extend this result to the principal component pursuit with reduced linear measurements, i.e. $Q^\perp$ is a $p$-dimensional random subspace. It's easy to note that results of paper \cite{Z12} is only a special case of ours.
\subsection{Basic problem formulations}
In this subsection, we will interpret an important strongly convex programming which will be addressed in this paper and list its existence and uniqueness theorems. In the paper \cite{A12}, the authors have studied principal component pursuit with reduced linear measurements and given sufficient conditions under which $L_0$ and $S_0$ can be perfectly recovered.
\begin{eqnarray}
\mbox{minimize}&& \|L\|_*+\lambda\|S\|_1 \nonumber\\
\mbox{subject to}&& \mathcal P_Q M=\mathcal P_Q (L+S)\nonumber
\end{eqnarray}

In this paper, we address a strongly convex programming. We prove it can guarantee exact low-rank matrix recovery. The proposed optimization is
\begin{eqnarray}
\mbox{minimize}&& \|L\|_*+\lambda\|S\|_1 +\frac{1}{2\tau}\|L\|_F^2+\frac{1}{2\tau}\|S\|_F^2\nonumber\\
\mbox{subject to}&& \mathcal P_Q M=\mathcal P_Q (L+S)
\end{eqnarray}
where $\tau\ge0$ is some positive penalty parameter and $P_Q$ is the orthogonal projection onto the linear subspace $Q$. We also assume $Q^\perp$ is a random subspace(the same assumption considered in paper \cite{A12}). When $\tau=\infty$ in (2), existence and uniqueness theorems is provided in the paper \cite{A12}, as we list them below. In the end, how to choose suitable parameters in the optimization model (2) is discussed.
\begin{theorem}\cite{A12} \label{T1}
       \emph{Fix any $C_p>0$, and let $Q^\perp$ be a $p$-dimensional random subspace
       of $\mathbb R^{n\times n}$; $L_0$ obeys incoherence condition with parameter $\mu$, and $\mbox{supp}(S_0)\sim \mbox{Ber}(\rho)$. Then with high probability, the solution of problem(2) with $\lambda=\frac{1}{\sqrt{n}}$ is exact, i.e. $\hat L = L_0$ and $\hat S = S_0$, provided that}
       \begin{eqnarray}
       \mbox{Rank}(L_0)< C_r n\mu^{-1}(\log{n})^{-2} ~~p<C_p n ~~
        \mbox{and} ~~\rho<\rho_0
       \end{eqnarray}
       \emph{where, $C_r$, $C_p$ and $\rho$ are positive numerical constants and $\rho_0<1$.}
\end{theorem}

\subsection{Contents and Notations}
We provide a brief summary of the notations which are used throughout the paper. We denote the operator norm of matrix by $\|X\| $, the Frobenius norm by $\|X\|_F $, the nuclear norm by  $\|X\|_* $, and the dual norm of $\|X\|_{(i)}$ by $\|X\|_{(i)}^*$. The Euclidean inner product between two matrices is defined by the formula $\left<X,Y\right>=trace(X^*Y)$. Note that $\|X\|_F^2 = \left<X,X\right>$. The Cauchy-Schwarz inequality gives $\left<X,Y\right>\le \|X\|_F \|Y\|_F$, and it is well known that we also have $\left<X,Y\right>\le \|X\|_{(i)} \|Y\|_{(i)}^*$, e.g.\cite{EB09} \cite{M01}. Linear transformations which act on the space of matrices are denoted by $\mathcal P X$. It's easy to see that the operator of $\mathcal P$ is high dimension matrix in substance. The operator norm of the operator is denoted by $\|\mathcal P\|$. It should be noted that $\|\mathcal P\|=\mbox{sup}_{\{\|X\|_F=1\}}\|\mathcal PX\|_F$.

The rest of the paper is organized as follows. In Section 2, we list many important Lemmas and prove a key lemma on which our main result depends. Suggestions then is given in Section 3, which will guide us to choose suitable parameters in practical algorithms. Conclusion and further works are discussed in Section 4.

\section{Important Lemmas}
In this section, we first list some useful lemmas which will be used throughout this paper and then prove a main lemma. Although the main lemma is similar to the corresponding one in the paper \cite{A12}, the construction of $W^Q$ is different. That leads to our necessary additional work.
\begin{lemma} [\cite{A12}, Lemma 1]\label{L1}
\emph{Suppose that $\mbox{dim}(Q^\perp\oplus T\oplus \Omega) =\mbox{dim}(Q^\perp) + \mbox{dim}(T) + \mbox{dim}(\Omega)$. Let $\Gamma=Q \cap T^\perp$ so that $\Gamma^\perp=Q^\perp \oplus T$. Assume that $\|\mathcal P_\Omega \mathcal P_{\Gamma^\perp} \| < 1/2$ and $\lambda <1$. Then, $(L_0, S_0)$ is the unique optimal solution to (2) if there exists a pair $(W,~F)\in\mathbb R^{n\times n} \times\mathbb R^{n\times n}$ satisfying}
\begin{eqnarray}
UV^*+W=\lambda(\mbox{sgn}(S_0)+F+\mathcal P_\Omega D)\in Q\nonumber
\end{eqnarray}
\emph{with $\mathcal P_T=0,~\|W\|<1/2,~\mathcal P_\Omega F=0,~\|F\|_\infty<1/2$, and $\|\mathcal P_\Omega D\|_F\le1/4$}.
\end{lemma}
\begin{lemma} [\cite{A12}, Lemma 3]\label{L2}
\emph{Assume that $\Omega \thicksim \mbox{Ber}(\rho)$ for some small $\rho\in(0,~1)$ and the other conditions of Theorem 1.2 hold true. Then, the matrix $W^L$ obeys, with high probability. \\}
$(a).~~ \|W^L\|<1/4$\\
$(b).~~ \|\mathcal P_\Omega(UV^*+W^L)\|_F<\lambda/4$\\
$(c).~~ \|\mathcal P_{\Omega^\perp}(UV^*+W^L)\|_\infty<\lambda/4$
\end{lemma}
\begin{lemma} [\cite{A12}, Lemma 4]\label{L3}
\emph{In addition to the assumptions in the previous lemma, assume that the signs of the non-zero entries of $S_0$ are i.i.d. random. Then, the matrix $W^S$ obeys, with high probability,\\}
$(a).~~ \|W^S\|<1/8$\\
$(b).~~ \|\mathcal P_{\Omega^\perp}W^S\|_\infty<\lambda/8$
\end{lemma}

The construction of $W^L$ and $W^S$ can be found in the paper \cite{A12}. The authors also introduce a new scheme to construct $W^Q$ for the principal component pursuit. However, the matrix $W^Q$ constructed in the paper \cite{A12} do not satisfy the requirement of our problem, so we have to modify this construction. We first give explicit construction of $W^Q$, and then, prove the modification of $W^Q$ satisfies the corresponding property.

\emph{Construction of $W^Q$ with least modification.} We define $W^Q$ by the following least squares problem:
\begin{eqnarray}
W^Q&=&\mbox{arg~min}_X~\|X\|_F\nonumber\\
\mbox{subject to} && \mathcal P_{Q^\perp}X=-\mathcal P_{Q^\perp}(UV^*+\frac{1}{\tau}L_0)\nonumber\\
&&\mathcal P_\Pi X=0\nonumber
\end{eqnarray}
where $\Pi= T\oplus \Omega$. This construction of $W^Q$ don't satisfy Theorem 2.6 only, but also has below Lemma.
\begin{lemma} \label{L4}
\emph{Assume $\tau\ge\|M\|_F$, and that $\Omega \thicksim \mbox{Ber}(\rho)$ for some small $\rho\in(0,~1)$ and the assumptions of Theorem 1.2 hold true. Then, the matrix $W^Q$ obeys, with high probability. \\}
$(a).~~ \|W^Q\|<1/8$\\
$(b).~~ \|\mathcal P_{\Omega^\perp}W^Q\|_\infty<\lambda/8$
\end{lemma}
In proof of Lemma 2.4, we have to use two important lemmas which are listed below.
\begin{lemma} [\cite{A12} Lemma 11] \label{L5}
  Let $S_1, S_2$ and $S_3$ be any three linear subspaces in $\mathbb R^{n\times n}$ satisfying $\mbox{dim}(S_1 \oplus S_2\oplus S_3) = \mbox{dim}(S_1)+\mbox{dim}(S_2)+\mbox{dim}(S_3)$, and $\mathcal P_{S_1}\mathcal P_{S_2}\le a_{1,2}< 1, \mathcal P_{S_2}\mathcal P_{S_3}\le a_{2,3}< 1$ and $\mathcal P_{S_3}\mathcal P_{S_1}\le a_{3,1}< 1$. We define $S = S_1 \oplus S_2$. Then, we have
\begin{eqnarray}
\|\mathcal P_S\mathcal P_{S_3}\|\le \sqrt{\frac{a_{2,3}^2+a_{3,1}^2}{1-a_{1,2}}}\nonumber
\end{eqnarray}
\end{lemma}
\begin{lemma} [\cite{A12} Lemma 7]
Assume that $p < n^2/4$. Let $Q^\perp$ be a linear subspace distributed according to the random subspace model. Then, with high probability, we have
\begin{eqnarray}
\|\mathcal P_{Q^\perp}\mathcal P_T\|\le 8 \frac{\sqrt{p}+\sqrt{2nr}}{n}\nonumber
\end{eqnarray}
\end{lemma}
\begin{proof}
\textbf{A, bounding the behavior of $\|UV^*+\frac{1}{\tau}L_0\|_F$}. For convenience, let $\xi:=\|UV^*+\frac{1}{\tau}L_0\|_F$.\\
According to triangle inequality, we have
\begin{eqnarray}
\|L_0\|_F=\|M-S_0\|_F\le\|M\|_F+\|S_0\|_F=\|M\|_F+\|\mathcal P_\Omega S_0\|_F\nonumber
\end{eqnarray}
In the last equality, we have used $S_0\in \Omega$. Note that
\begin{eqnarray}
\|\mathcal P_\Omega S_0\|_F=\|\mathcal P_\Omega (M-L_0)\|_F\le\|\mathcal P_\Omega M\|_F+\|\mathcal P_\Omega L_0\|_F\nonumber
\end{eqnarray}
According to the derivation in the paper \cite{Z12}, with high probability, we can obtain
\begin{eqnarray}
\|\mathcal P_\Omega L_0\|_F\le\frac{\sqrt{3}}{3}\|\mathcal P_{\Omega^\perp} M\|_F\le\frac{\sqrt{3}}{3}\| M\|_F\nonumber
\end{eqnarray}
Putting those all together, we get
\begin{eqnarray}
\|L_0\|_F\le(\sqrt{3}/3+2)\|M\|_F\nonumber
\end{eqnarray}
Combining with $\tau\ge\|M\|_F$, we can obtain
\begin{eqnarray}
\xi\le\|UV^*\|_F+\frac{\|L_0\|_F}{\tau}\le r+\frac{(\sqrt{3}/3+2)\|M\|_F}{\tau}\le r+\sqrt{3}/3+2\nonumber
\end{eqnarray}
Because $W^Q$ is the optimum solution of least squares problem, we can use the convergent Neumann series expansion. It's easy to note that
\begin{eqnarray}
W^Q=\mathcal P_{\Pi^\perp}\sum_{k>0}(\mathcal P_{Q^\perp}\mathcal P_\Pi\mathcal P_{Q^\perp})^k(\mathcal P_{Q^\perp}(-UV^*-\frac{1}{\tau}L_0))\nonumber
\end{eqnarray}
According to triangle inequality, we have
\begin{eqnarray}
\|W^Q\|_F\le\|\sum_{k>0}(\mathcal P_{Q^\perp}\mathcal P_\Pi\mathcal P_{Q^\perp})^k\|\|\mathcal P_{Q^\perp}(-UV^*-\frac{1}{\tau}L_0)\|_F
\end{eqnarray}
\textbf{B, estimating the first inequality of Lemma 2.4}.
In order to bound $\|W^Q\|_F$, we have to bound the behavior of $\|\sum_{k>0}(\mathcal P_{Q^\perp}\mathcal P_\Pi\mathcal P_{Q^\perp})^k\|$. Therefore, we have
\begin{eqnarray}
\|\sum_{k>0}(\mathcal P_{Q^\perp}\mathcal P_\Pi\mathcal P_{Q^\perp})^k\|&\le&\sum_{k>0}\|(\mathcal P_{Q^\perp}\mathcal P_\Pi\mathcal P_{Q^\perp})^k\|\nonumber\\
&\le&\sum_{k>0}\|\mathcal P_{Q^\perp}\mathcal P_\Pi\|^{2k}\nonumber
\end{eqnarray}
According to Lemma 2.5, we have, for any $\epsilon>0$, with high probability,
\begin{eqnarray}
\|\mathcal P_{Q^\perp}\mathcal P_\Pi\|^{2}\le\frac{64}{1-\sqrt{\rho+\epsilon}}
\left(
\left(
\sqrt{\frac{p}{n^2}}+\sqrt{\frac{5\rho}{4}}
\right)^2
+
\left(
\sqrt{\frac{p}{n^2}}+\sqrt{\frac{2r}{n}}
\right)^2
\right)\nonumber
\end{eqnarray}
According to the paper\cite{A12}, we have
\begin{eqnarray}
\|\sum_{k>0}(\mathcal P_{Q^\perp}\mathcal P_\Pi\mathcal P_{Q^\perp})^k\|\le\frac{4}{3}\nonumber
\end{eqnarray}
with high probability.

Next, we will bound $\|\mathcal P_{Q^\perp}(-UV^*-\frac{1}{\tau}L_0)\|_F$. According to the paper \cite{A12}, $P_{Q^\perp}$ has the same distribution as $H(H^*H)^{-1}H^*$, where $H \in\mathbb R^{n^2\times p}$ is a random Gaussian matrix with i.i.d. entries $\sim \mathcal N(0, 1/n^2)$. Therefore, we can obtain
\begin{eqnarray}
&&\|\mathcal P_{Q^\perp}(UV^*+\frac{1}{\tau}L_0)\|_F\nonumber\\
&=&\|H(H^*H)^{-1}H^*\mbox{vec}(UV^*+\frac{1}{\tau}L_0)\|_F\nonumber\\
&\le&\|H(H^*H)^{-1}\|\|H^*\mbox{vec}(UV^*+\frac{1}{\tau}L_0)\|_2\nonumber
\end{eqnarray}
Together with Lemma 2.6, we can obtain
\begin{eqnarray}
\mathbb P[\|H(H^*H)^{-1}\|\ge4]\le e^{-\frac{n^2}{32}}\nonumber
\end{eqnarray}
It's easy to note that any entries of $H^*\mbox{vec}(UV^*+\frac{1}{\tau}L_0)$ have the same distribution as $<G,~UV^*+\frac{1}{\tau}L_0>$, where $G_{ij}\sim\mathcal N(0, 1/n^2)$ are independent identically distributed. It is obvious to see that
\begin{eqnarray}
\mathbb E\{<G,~UV^*+\frac{1}{\tau}L_0>\}=<\mathbb E\{G\},~UV^*+\frac{1}{\tau}L_0>=0\nonumber
\end{eqnarray}
and
\begin{eqnarray}
\mbox{Var}\{<G,~UV^*+\frac{1}{\tau}L_0>\}&=&\sum_{ij}(UV^*+\frac{1}{\tau}L_0)_{ij}^2\mbox{Var}\{G_{ij}\}\nonumber\\
&=&\xi^2/n^4\nonumber
\end{eqnarray}
 Therefore, $<G,~UV^*+\frac{1}{\tau}L_0>$ is distributed according to $\sim\mathcal N(0, \xi/n^2)$, where $\xi:=\|UV^*+\frac{1}{\tau}L_0\|_F$. For simplicity, we define $Z:=H^*\mbox{vec}(UV^*+\frac{1}{\tau}L_0)$. Using the Jesen inequality, we have
\begin{eqnarray}
\mathbb E[\|Z\|_2]\le (\mathbb E[\|Z\|_2^2])^{1/2}=\sqrt{\frac{p\xi}{n^2}}\nonumber
\end{eqnarray}
According to the Proposition 2.18 in \cite{M01}, we can obtain
\begin{eqnarray}
\mathbb P\left[\|Z\|_2\ge \mathbb E[\|Z\|_2]+ t\sqrt{\frac{\xi}{n^2}}\right] \le e^{-t^2/2}\nonumber
\end{eqnarray}
Setting $t=\sqrt{6\mbox{log}n}$, after a simple calculation, we can obtain
\begin{eqnarray}
\|W^Q\|\le \|W^Q\|_F\le\frac{16}{3}\left( \frac{\sqrt{p\xi}}{n}+ \frac{\sqrt{6\xi\mbox{log}n}}{n}\right)\nonumber
\end{eqnarray}
with high probability. For sufficiently large $n$, the first inequality of Lemma 2.4 is established.\\
\textbf{C, estimating the second inequality of Lemma 2.4},
Note that
\begin{eqnarray}
W^Q=\mathcal P_{\Pi^\perp}\mathcal P_{Q^\perp}\sum_{k>0}(\mathcal P_{Q^\perp}\mathcal P_\Pi\mathcal P_{Q^\perp})^k(\mathcal P_{Q^\perp}(-UV^*-\frac{1}{\tau}L_0))\nonumber
\end{eqnarray}
Similar to the paper \cite{A12}, after a simple calculation, we can obtain
\begin{eqnarray}
\|\mathcal P_{\Omega^\perp}W^Q\|_\infty\le \frac{C\sqrt{\xi}}{n^2}(\sqrt{p}+\sqrt{6\log{n}})^2\nonumber
\end{eqnarray}
where $C$ is some constant. Note that for sufficiently large $n$, the second inequality of Lemma 2.4 is established.
\end{proof}

\section{Bounding parameter $\tau$}
In this section, we shall provide sufficient conditions under which $(L_0; S_0)$ is the unique solution of the strongly convex programming (2) with high probability. Afterwards, an explicit lower bound of $\tau$ will be given as well, which will guide us to choose suitable parameters in practical algorithms.
\begin{theorem} \label{result1}
\emph{Suppose that $\mbox{dim}(Q^\perp\oplus T\oplus \Omega) =\mbox{dim}(Q^\perp) + \mbox{dim}(T) + \mbox{dim}(\Omega)$. Let $\Gamma=Q \cap T^\perp$ so that $\Gamma^\perp=Q^\perp \oplus T$. Assume that $\|\mathcal P_\Omega \mathcal P_{\Gamma^\perp} \| < 1/2$ and $\lambda <1$. If there exists a pair $(W,~F)\in\mathbb R^{n\times n} \times\mathbb R^{n\times n}$ and a matrix $D$ satisfying}
\begin{eqnarray}
UV^*+W+\frac{1}{\tau}L_0=\lambda(\mbox{sgn}(S_0)+F+\mathcal P_\Omega D)+\frac{1}{\tau}S_0\in Q\nonumber
\end{eqnarray}
\emph{with}
\begin{eqnarray}
\mathcal P_TW=0,~\|W\|\le \beta,~\mathcal P_\Omega F=0,~\|F\|_\infty\le\beta,~ \|\mathcal P_\Omega D\|_F\le\alpha
\end{eqnarray}
\emph{where $\alpha$, $\beta$ are positive parameters satisfying}
\begin{eqnarray}
 \alpha+\beta\le1
\end{eqnarray}
\emph{Then $(L_0, S_0)$ is the unique solution of the strongly convex programming (2)}.
\end{theorem}
\begin{proof}
For any feasible perturbation $(H_L,~H_S)$, it's easy to note that $\mathcal P_QH_L=\mathcal P_QH_S$. According to the definition of $\Gamma$, we have $\Gamma\subset Q$, therefore $\mathcal P_\Gamma H_L=\mathcal P_\Gamma H_S$. For simplicity, let $f(L,S)=\|L\|_*+\lambda\|S\|_1+\frac{1}{2\tau}\|L\|_F^2+\frac{1}{2\tau}\|S\|_F^2$, we can obtain
\begin{eqnarray}
&&f(L_0+H_L, S_0-H_S)\nonumber\\
& \ge& f(L_0,S_0)+ <UV^*+W_0+\frac{1}{\tau}L_0, H_L>-<\lambda \mbox{sgn}(S_0)+\lambda F_0+\frac{1}{\tau}S_0, H_S>\nonumber\\
& \ge& f(L_0,S_0)+ <W_0, H_L> -<W, H_L>+ <UV^*+W+\frac{1}{\tau}L_0, \mathcal P_Q H_L>\nonumber\\
&&~~-<\lambda F_0, H_S> +<\lambda F, H_S> -<\lambda \mbox{sgn}(S_0)+\lambda F+\frac{1}{\tau}S_0, \mathcal P_Q H_S>\nonumber\\
& \ge& f(L_0,S_0)+ <W_0, \mathcal P_{T^\perp}H_L> -<W, \mathcal P_{T^\perp}H_L>\nonumber\\
&&~~-<\lambda F_0, \mathcal P_{\Omega^\perp}H_S> +<\lambda F, \mathcal P_{\Omega^\perp}H_S> -<\lambda \mathcal P_\Omega D, \mathcal P_Q H_S>\nonumber\\
& \ge& f(L_0,S_0)+ (1-\beta)\|\mathcal P_{T^\perp}H_L\|_* + (1-\beta)\lambda\|\mathcal P_{\Omega^\perp}H_S\|_1-\alpha\lambda\|\mathcal P_\Omega H_S\|_F\nonumber
\end{eqnarray}
In the second inequality above, we have used the facts
\begin{eqnarray}
UV^*+W+\frac{1}{\tau}L_0=\lambda (\mbox{sgn}(S_0)+ F+\mathcal P_Q D)+\frac{1}{\tau}S_0\in Q\nonumber
\end{eqnarray}
In the third inequality above, we have used $\mathcal P_QH_L=\mathcal P_QH_S$.\\
We will bound $\|\mathcal P_\Omega H_S\|_F$. According to the definition of $\Gamma$ , we have
\begin{eqnarray}
\|\mathcal P_\Omega H_S\|_F&\le&\|\mathcal P_\Omega\mathcal P_\Gamma H_S\|_F+\|\mathcal P_\Omega \mathcal P_{\Gamma^\perp} H_S\|_F\nonumber\\
&\le&\|\mathcal P_\Omega\mathcal P_\Gamma H_L\|_F+\frac{1}{2}\| H_S\|_F\nonumber\\
&\le&\|\mathcal P_\Gamma H_L\|_F+\frac{1}{2}\|\mathcal P_\Omega H_S\|_F+\frac{1}{2}\| \mathcal P_{\Omega^\perp} H_S\|_F\nonumber\\
&\le&\|\mathcal P_{T^\perp} H_L\|_F+\frac{1}{2}\|\mathcal P_\Omega H_S\|_F+\frac{1}{2}\| \mathcal P_{\Omega^\perp} H_S\|_F\nonumber
\end{eqnarray}
Therefore
\begin{eqnarray}
\|\mathcal P_\Omega H_S\|_F\le 2\|\mathcal P_{T^\perp} H_L\|_F+ \|\mathcal P_{\Omega^\perp} H_S\|_F \le2\|\mathcal P_{T^\perp} H_L\|_*+ \|\mathcal P_{\Omega^\perp} H_S\|_1\nonumber
\end{eqnarray}
Putting those all together, we get
\begin{eqnarray}
&&f(L_0+H_L, S_0-H_S)\nonumber\\
& \ge& f(L_0,S_0)+ (1-\beta-2\alpha\lambda)\|\mathcal P_{T^\perp}H_L\|_* + (1-\beta-\alpha)\lambda\|\mathcal P_{\Omega^\perp}H_S\|_1\nonumber
\end{eqnarray}
This, together with (6), implies that $(L_0,~S_0)$ is a solution to (2). The uniqueness follows from the strong convexity of the objective in (2).
\end{proof}
We will provide the criterion of the value of $\tau$.
\begin{theorem} \label{result1}
\emph{Let $\tau_1=\frac{\|\mathcal P_{\Omega^\perp}L_0\|_\infty}{(\beta-\frac{1}{2})\lambda}$, $\tau_2=\frac{\|\mathcal P_\Omega(L_0-S_0)\|_F}{(\alpha-\frac{1}{4})\lambda}$, and $\tau_3=\frac{4(\|\mathcal P_{\Omega^\perp}L_0\|_\infty+\|\mathcal P_\Omega(L_0-S_0)\|_F)}{\lambda}$. Assume}
\begin{eqnarray}
 \tau\ge\mbox{max}\left(\tau_1,~\tau_2,
 ~\tau_3,\|M\|_F\right)
\end{eqnarray}
Then, under the other assumptions of Theorem 1.1, $(L_0, S_0)$ is the unique solution to the strongly convex programming (2) with high probability.
\end{theorem}
\begin{proof}
In order to check the conditions in Theorem 3.1, we will prove there exists a matrix $W$ obeying
\begin{eqnarray}
 \left\{
 \begin{array}{ll}
  \mathcal P_T W=0\\
  \|W\|\le\beta\\
  \mathcal P_{Q^\perp}W=-\mathcal P_{Q^\perp}(UV^*+\frac{1}{\tau}L_0)\\
  \|\mathcal P_{\Omega^\perp}(UV^*+W+\frac{1}{\tau}L_0-\frac{1}{\tau}S_0)\|_\infty\le\beta\lambda\\
  \|\mathcal P_{\Omega}(UV^*+W-\lambda \mbox{sgn}(S_0)+\frac{1}{\tau}L_0-\frac{1}{\tau}S_0)\|_F\le\alpha\lambda\\
       \end{array}
    \right.
\end{eqnarray}
 Note that $W = W^L +W^S+W^Q$ with $W^L$, $W^S$ and $W^Q$ have analytical form constructed in the paper \cite{A12}. We will check above conditions hold true one by one. For simplicity of proof, we denote
\begin{eqnarray}
\gamma:=\|\mathcal P_{\Omega^\perp}(L_0-S_0)\|_\infty, ~~~\delta:=\|\mathcal P_\Omega(L_0-S_0)\|_F\nonumber
\end{eqnarray}
Without loss of generality, let $\beta > 1/2$. With the help of the construction \cite{A12} of $W^L$, $W^S$ and $W^Q$, it is easy to check the first and second conditions hold true. With respect to the third condition, according to the paper \cite{A12}, we have $\mathcal P_{Q^\perp}W^L=0$ and $\mathcal P_{Q^\perp}W^S=0$. According to the modification of $W^Q$ constructed in Lemma 2.4, we have $\mathcal P_{Q^\perp}W^Q=-\mathcal P_{Q^\perp}(UV^*+\frac{1}{\tau}L_0)$. It's easy to check that $\mathcal P_{Q^\perp}W=\mathcal P_{Q^\perp}W^L+\mathcal P_{Q^\perp}W^S+\mathcal P_{Q^\perp}W^Q=-\mathcal P_{Q^\perp}(UV^*+\frac{1}{\tau}L_0)$, which implies that the third condition holds true. Consequently, we will provide the last two conditions also hold true under some suitable assumptions. Pertaining to the fourth inequality, we have
\begin{eqnarray}
&&\|\mathcal P_{\Omega^\perp}(UV^*+W+\frac{1}{\tau}L_0-\frac{1}{\tau}S_0)\|_\infty\nonumber\\
&\le& \|\mathcal P_{\Omega^\perp}(UV^*+W^L)\|_\infty+\|\mathcal P_{\Omega^\perp}W^S\|_\infty\nonumber\\
&&~~+\|\mathcal P_{\Omega^\perp}W^Q\|_\infty+\frac{1}{\tau}\|\mathcal P_{\Omega^\perp}(L_0-S_0)\|_\infty\nonumber\\
&\le& \frac{\lambda}{4}+\frac{\lambda}{8}+\frac{\lambda}{8}+\frac{1}{\tau}\|\mathcal P_{\Omega^\perp}(L_0-S_0)\|_\infty\nonumber\\
&\le& \frac{\lambda}{2}+\frac{\gamma}{\tau}\nonumber
\end{eqnarray}
For the last inequality, noting that $\mathcal P_\Omega(W^S) = \lambda \mbox{sgn}(S_0)$ and $\mathcal P_\Omega (W^Q)=0$ as shown in \cite{A12}, we can obtain
\begin{eqnarray}
&&\|\mathcal P_\Omega(UV^*+W-\lambda\mbox{sgn}(S_0)+\frac{1}{\tau}L_0-\frac{1}{\tau}S_0)\|_F\nonumber\\
&=&\|\mathcal P_{\Omega}(UV^*+W^L+\frac{1}{\tau}L_0-\frac{1}{\tau}S_0)\|_F\nonumber\\
&\le& \|\mathcal P_{\Omega}(UV^*+W^L)\|_F+\frac{1}{\tau}\|\mathcal P_{\Omega}(L_0-S_0)\|_F\nonumber\\
&\le& \frac{\lambda}{4}+\frac{\delta}{\tau}\nonumber
\end{eqnarray}
In order to satisfy the condition (8), we choose a $\tau$ obeying
\begin{eqnarray}
\frac{\lambda}{2}+\frac{\gamma}{\tau}\le\beta\lambda,~~\mbox{and}~~\frac{\lambda}{4}
+\frac{\delta}{\tau}\le\alpha\lambda
\end{eqnarray}
Therefore
\begin{eqnarray}
\tau\ge\mbox{max}\left(\frac{\gamma}{(\beta-\frac{1}{2})\lambda},~\frac{\delta}{(\alpha-\frac{1}{4})\lambda}\right)
\end{eqnarray}
Combining (9) with (6), we can obtain
\begin{eqnarray}
\frac{\lambda}{2}+\frac{\gamma}{\tau}+\frac{\lambda}{4}
+\frac{\delta}{\tau}\le\beta\lambda+\alpha\lambda\le\lambda\nonumber
\end{eqnarray}
Therefore
\begin{eqnarray}
\tau\ge\frac{4(\gamma+\delta)}{\lambda}
\end{eqnarray}
Together with (10) and (11), the Theorem 3.2 is established.
\end{proof}
In order to simplify the formula (7), we suppose $\alpha=3/8$ and $\beta=5/8$, which satisfy the conditions above. Therefore
 \begin{eqnarray}
 \tau\ge\mbox{max}\left(\frac{8\|\mathcal P_{\Omega^\perp}L_0\|_\infty}{\lambda},~\frac{8\|\mathcal P_\Omega(L_0-S_0)\|_F}{\lambda}\right)
\end{eqnarray}
However, note that the exact lower bound is very hard to get, because we only have the information about the given data matrix $M$. Noting that
\begin{eqnarray}
\|\mathcal P_{\Omega^\perp}M\|_\infty\le\|M\|_\infty\nonumber
\end{eqnarray}
 And according to the paper \cite{Z12}, we have
\begin{eqnarray}
\|\mathcal P_{\Omega}(L_0-S_0)\|_F&\le&\frac{\sqrt{15}}{3}\|M\|_F\nonumber
\end{eqnarray}
Therefore, we can choose
 \begin{eqnarray}
 \tau\ge\mbox{max}\left(\frac{8\|M\|_\infty}{\lambda},~\frac{8\sqrt{15}\|M\|_F}{3\lambda}\right)\nonumber
\end{eqnarray}
It's obvious that $\|M\|_\infty\le\|M\|_F$. Therefore, we can obtain the result as follows.
\begin{theorem} \label{Tau2}
        \emph{~~Assuming }
         \begin{eqnarray}
         \tau\ge~\frac{8\sqrt{15}\|M\|_F}{3\lambda}\nonumber
         \end{eqnarray}
        \emph{and the other assumptions of Theorem 1.1, $(L_0, S_0)$ is the unique solution to the strongly convex programming (2) with high probability.}
\end{theorem}

\section{Conclusion}
In this paper, we have studied strongly convex programming for principal component pursuit with reduced linear measurements. We first provide sufficient conditions under which the strongly convex models lead to the exact low rank and sparse components recovery; Second, we give the criterion of the choice of the value of $\tau$, which gives very useful advice on how to set the suitable parameters in designing efficient algorithms. Especially, it is easy to note that the main results of paper \cite{Z12} is only the special case of ours. In some sense, We extend the result of choosing suitable parameters to the general problem.

\section*{Acknowledgements}
We would like to thank the reviewers very much for their valuable comments and suggestions. This research was supported by the National Natural Science Foundation of China (NSFC) under Grant 61172140, and '985' key projects for excellent teaching team supporting (postgraduate) under Grant A1098522-02. Yipeng Liu is supported by FWO PhD/postdoc grant:G0108.11(compressed sensing).

\medskip



\medskip
 {\it E-mail address: }youlin\_2001@163.com\\
 \indent{\it E-mail address: }wanqun@uestc.edu.cn\\
  \indent{\it E-mail address: }yipeng.liu@esat.kuleuven.be\\
\end{document}